\documentclass[twoside,11pt]{article}

\usepackage{amssymb}
\usepackage{amsmath}
\usepackage{amsfonts}

\begin{document}           

\title{\large\textsc{THE BIG BANG QUANTUM COSMOLOGY:\\ THE MATTER-ENERGY PRODUCTION EPOCH}}
\author{\textsc{V.E. Kuzmichev, V.V. Kuzmichev}\\[0.5cm]
 Bogolyubov Institute for Theoretical Physics,\\
 National Academy of Sciences of Ukraine,\\ 
 Kiev, 03680 Ukraine}

\date{}

\maketitle

\setcounter{page}{1}

\begin{abstract}
The exactly solvable quantum model of the homogeneous, isotropic and closed
universe in the matter-energy production epoch is considered. It is assumed that the
universe is originally filled with a uniform scalar field and a perfect fluid which
defines a reference frame. The stationary state spectrum and the wave functions of 
the quantum universe are calculated. In this model the matter-energy in the 
universe has a component in the form of a condensate of massive zero-momentum 
excitation quanta of oscillations of primordial 
scalar field. The mean value of the scale factor of the universe in a given state is 
connected with the mass of a condensate by a linear relation. 
The nucleation rate of the 
universe from the initial cosmological singularity point is calculated. It is 
demonstrated that the process of nucleation of the universe can have an exponential 
(explosive) nature. 
The evolution of the universe is described as 
transitions with non-zero probabilities between the states of the 
universe with different masses of a condensate.
\end{abstract}

PACS numbers: 98.80.Qc, 04.60.-m, 04.60.Kz 

\begin{center}
      \textbf{1. Introduction}\\[0.5cm]
\end{center}
The method of constraint system quantization can be taken as a basis of quantum 
theory of gravity suitable for the investigation of cosmological systems \cite{Dir}. 
As is well known the structure of constraints in general relativity is such that
variables which correspond to true dynamical degrees of freedom cannot be 
singled out from canonical variables of geometrodynamics. The reason behind this 
difficulty is the absence of predetermined way of identifying spacetime events in 
generally covariant theory \cite{KuchT}.

In contrast to gravitational field in a void, the 
consideration of gravitational field
coupled with matter allows to use matter in order to give an invariant meaning to 
space-time points \cite{KuchT,BroM,Tor}. 
Using material reference systems, one can address 
conceptual problems of not only classical, but also quantum gravity \cite{DeW}. 

In Ref. \cite{KuchT} a scheme to include reference frames in general relativity by
means of an introduction of coordinate conditions was developed. A task at this 
point is to find 
a material source in the Einstein equations which determines a reference 
frame and has no unphysical properties. This approach was applied in Refs. 
\cite{Kuz1,KuzK} in order to solve the problems of quantum theory of 
gravity in the minisuperspace model with a material source in the form of
relativistic matter (radiation) which defines the reference frame. The variables 
which describe radiation mark spacetime events, since the reference frame is
considered as a dynamical system. These variables play the role of the canonical 
coordinates which determine an embedding in the encompassing spacetime. At the same 
time the new constraints turn out to be linear with respect to the momenta 
canonically conjugate with them. Such an approach allows to obtain the time-dependent 
Schr\"{o}dinger equation and to define a conserved positive definite inner product.

On the other hand 
it is well known that during the quantization of different model systems
in gravity one can use a perfect fluid as a reference frame \cite{BroM,Lun,Dem}. In
this case one deals directly with a physical medium without  
coordinate conditions as an intermediate under the construction of a material 
source. This leaves aside problems connected with the necessity to ensure that 
the action is coordinate invariant and that a material source which
determines a reference frame has correct physical properties. 
Relativistic matter is a special case of a perfect
fluid and as the simplest physical system can be used to define a reference frame.

In Section~2 the summary of ideas that lead to the fundamental equations
of quantum model of the universe is given.
   
In Section~3 the stationary state spectrum and the wave functions of the 
quantum universe filled with primordial matter in the form of a uniform scalar 
field and a perfect fluid (radiation) which defines a reference frame are 
calculated. It is shown that the matter-energy in the universe has a component in the
form of a condensate of massive zero-momentum excitation quanta of oscillations of 
a primordial scalar field. The mean value of the scale factor in a given state of the 
universe is calculated. This mean value is determined by the mass of a condensate and 
the mean deviation from an equilibrium state that can be neglected in 
the semi-classical limit. It is shown that the universe with the Planck mass of a
condensate in the ground (vacuum) state has the mean value of the scale factor that
practically coincides with the Planck length. 

In Section~4 the nucleation rate of the universe from the initial cosmological 
singularity point is calculated and it is demonstrated that the process of 
nucleation of the universe can have an exponential (explosive) nature. This 
phenomenon can be identified with the initial moment of the Big Bang. It is shown 
that the greater masses of a condensate of the nucleating universe correspond to 
its larger initial sizes.

In Section~5 the probabilities of transitions between the states of the 
universe with different masses of a condensate are calculated. It is shown that the
probability of transition from the vacuum state of the universe to another state 
obeys the Poisson distribution.

In Section~6 a conclusion is given.

\begin{center} 
      \textbf{2. Constraint system quantization in the presence
       of a medium which defines a reference frame }\\[0.5cm]
\end{center}

Let us consider a cosmological system (universe) with the action 
\begin{equation}\label{01}
    S = S_{E-H} + S_{M} + S_{PF},
\end{equation}
where
\begin{equation}\label{02}
   S_{E-H} = - \frac{c^{3}}{16 \pi G} \int\! d^{4}x \sqrt{-\, g}\, R
\end{equation}
is the Einstein-Hilbert action for gravitational field, $S_{M}$ is the action
of matter fields, 
\begin{eqnarray}\label{03}
    S_{PF} = \frac{1}{c}\, \int\! d^{4}x \sqrt{-\, g}\,\{ -\, \rho(\rho_{0}, s)\, +\, 
\lambda [g^{\mu\nu}U_{\mu}U_{\nu} - 1 ]\nonumber \\
 +\, \rho_{0} U^{\nu} [s \,\Theta_{,\,\nu} \,-\, \widetilde{\lambda}_{,\,\nu}\, +\, 
\beta_{i}\,\alpha_{,\,\nu}^{i}]\,  \}
\end{eqnarray}
is the action of a perfect fluid 
(macroscopic bodies) \cite{Taub,SeW,Sch70,Br92,Kuz4}, which defines material
reference frame, where $\rho$ is the energy density as a function of the density
of the rest mass $\rho_{0}$ and the specific entropy $s$, $U^{\nu}$ is the 
four-velocity, $\lambda$ is a Lagrange multiplier that ensures normalization of
$U^{\nu}$. The $\Theta,\, \widetilde{\lambda},\, \beta_{i},\, \alpha^{i}$ are scalar
fields. Here $\Theta$ has a meaning of the thermasy or the ``potential'' for the 
temperature $T$, $T = \Theta_{,\, \nu} U^{\nu}$. The $\widetilde{\lambda}$ is the 
``potential'' for the specific free energy $f$ taken with an inverse sign, $f = -\, 
\widetilde{\lambda}_{, \, \nu} U^{\nu}$. 
The $\beta_{i}$ and $\alpha^{i}$ are Lagrange 
multipliers and Lagrangian coordinates for a fluid on a spacelike hypersurface 
respectively (one should introduce them into the action in order to describe 
rotational flows and, generally speaking, 
an incorporation of one pair of the variables 
$\beta_{i}$ and $\alpha^{i}$ into the action would be enough, but then their direct 
physical interpretation will be lost). The thermodynamic variables are related 
via the first law of thermodynamics
\begin{equation}\label{04}
    d\rho = h\, d\rho_{0}\, +\, \rho_{0}\, T\, ds,
\end{equation}
where $h = \frac{\rho + p}{\rho_{0}}$ is the specific enthalpy which plays the role
of inertial mass, $p$ is the pressure.

The components of the metric tensor $g^{\mu\nu}$, the matter fields contained in
$S_{M}$, and the values $\rho_{0},\, s,\, U^{\nu},\, \Theta,\, \lambda,\, 
\widetilde{\lambda},\, \beta_{i}$ and $\alpha^{i}$ play the role of generalized
variables. All equations of classical theory of gravity follow from the principle of
least action
\begin{equation}\label{05}
    \delta S = 0,
\end{equation}
where all independent variables are varied \cite{Kuz4}.

Let us assume that the universe is homogeneous, isotropic, closed and described by 
the Robertson-Walker metric
\begin{equation}\label{1}
    ds^{2} =  N^{2}(t) c^{2} dt^{2} -  a^{2}(t) d\Omega^{2},
\end{equation}
where $a(t)$ is the cosmic scale factor, $N(t)$ is the lapse function that 
specifies the time reference scale,  $t$ is the time variable, $d\Omega^{2}$ is an 
interval element on a unit three-sphere. We choose the uniform scalar field $\phi$
with the potential energy density (potential) $V(\phi)$ as a matter. The choice of 
such a field as a primordial matter seems to be reasonable, since any other
fields (vector or spinor, for instance), being non-averaged over all space variables 
at every instant of time, can destroy the supposed property 
of homogeneity and isotropy of the universe.

In the model of the universe under consideration the action (\ref{01}) can be 
reduced to the form \cite{Kuz4}
\begin{equation}\label{06}
    S = \int\! dt\,\left(\pi_{a} \frac{da}{dt} + \pi_{\phi} \frac{d\phi}{dt} +  
\pi_{\Theta} \frac{d\Theta}{dt} + \pi_{\widetilde{\lambda}} 
\frac{d\widetilde{\lambda}}{dt} -  H \right),
\end{equation}
where $\pi_{a},\,\pi_{\phi},\, \pi_{\Theta},\, \pi_{\widetilde{\lambda}}$ are
the momenta canonically conjugate with the variables $a,\, \phi,\, \Theta,\, 
\widetilde{\lambda}$, and it is taken into account that the momenta conjugate with
the variables $\rho_{0},\, s$ and $N$, vanish identically. In this model $U^{0} = 
\frac{1}{N},\, U^{i} = 0$, the condition $g^{\mu\nu}U_{\mu}U_{\nu} = 1$ is
contained explicitly in the variational principle and, therefore, the term with 
$\lambda$ should be dropped. Since the variables $\beta_{i}$ and $\alpha^{i}$ 
describe rotational flows which would create a preferential direction in space,
the terms with $\beta_{i}$ and $\alpha^{i}$ should be dropped as well and one should 
consider irrotational (potential) flows of a perfect fluid only.

The $H$ in Eq. (\ref{06}) is the Hamiltonian
\begin{eqnarray}\label{07}
    H = N \left(-\, \frac{3 \pi c^{4}}{4 G}\right)\,\frac{1}{a} \left\{\left(\frac{2
 G}{3 \pi c^{3}} \right)^{2} \pi_{a}^{2} +  a^{2} - \frac{G}{3 \pi^{3} c^{2}}\,\, 
\frac{\pi_{\phi}^{2}}{a^{2}} - \frac{8 \pi G}{3 c^{4}}\, a^{4} [\rho + 
V(\phi)]\right\} \nonumber \\ 
+ \lambda_{1}\left\{\pi_{\Theta} - \frac{1}{c}\,2\pi^{2}  a^{3} \rho_{0}\, s\right\}
  + \lambda_{2}\left\{\pi_{\widetilde{\lambda}} + \frac{1}{c}\,2\pi^{2}  a^{3} 
\rho_{0} \right\},
\end{eqnarray}
where $\lambda_{1}$ and $\lambda_{2}$ are Lagrange multipliers. 
The function $N$ in (\ref{07}) plays also the role of a Lagrange multiplier (like in
ADM-formalism \cite{ADM}). The variation of the action (\ref{06}) with respect to
$N,\, \lambda_{1}$ and $\lambda_{2}$ leads to three constraint equations.

Taking into account the conservation laws (see below)
and vanishing of the momenta conjugate with $\rho_{0}$ 
and $s$, one can discard degrees of freedom corresponding to these variables
and convert the second-class constraints into first-class constraints 
in accordance with Dirac's proposal \cite{Dir}. In quantum theory 
first-class constraint equations become constraints on the wave function $\Psi$. 

Replacing the momenta by the operators
\begin{eqnarray*}
   \pi_{a} \rightarrow \hat{\pi}_{a} & = & -\, i\,\hbar \partial_{a},\quad
   \pi_{\phi} \rightarrow \hat{\pi}_{\phi} = -\, i\,\hbar \partial_{\phi}, \\
\pi_{\Theta} \rightarrow \hat{\pi}_{\Theta} & = & -\, i\,\hbar \partial_{\Theta}, \quad
\pi_{\widetilde{\lambda}} \rightarrow \hat{\pi}_{\widetilde{\lambda}} = -\, i\,\hbar 
\partial_{\widetilde{\lambda}}, 
\end{eqnarray*}
which satisfy the commutation relations
$$[a, \hat{\pi}_{a}] = i\,\hbar, \quad [\phi, \hat{\pi}_{\phi}] = i\,\hbar,\quad 
[\Theta, \hat{\pi}_{\Theta}] = i\,\hbar,\quad 
[\widetilde{\lambda}, \hat{\pi}_{\widetilde{\lambda}}] = i\,\hbar, $$
while all other commutators vanish, we obtain
\begin{equation}\label{08}
    \left\{-\, \left(\frac{2 G \hbar}{3 \pi c^{3}} \right)^{2}\partial^{2}_{a}\, +\, 
a^{2}\, +\, \frac{G \hbar^{2}}{3 \pi^{3} c^{2}}\,\, \frac{1}{a^{2}}\,\, 
\partial^{2}_{\phi}\, -\,\frac{8 \pi G}{3 c^{4}}\,\, a^{4} [\rho\, +\, 
V(\phi)]\right\} \Psi = 0, 
\end{equation}
\begin{equation}\label{09}
    \left\{-\, i\,\hbar \,\partial_{\Theta} - \frac{1}{c}\, 2 \pi^{2} a^{3} 
\rho_{0}\, s\right\} \Psi = 0,
\end{equation}
\begin{equation}\label{010}
    \left\{-\, i\,\hbar \,\partial_{\widetilde{\lambda}} + \frac{1}{c}\, 2 \pi^{2} 
a^{3} \rho_{0} \right\} \Psi = 0.
\end{equation}
In Eq. (\ref{08}) the factor ordering parameter associated with a possible ambiguity 
in the choice of an explicit form of the operator $\hat{\pi}_{a}^{2}$ 
is assumed to be zero.

It is convenient to pass from the generalized variables $\Theta$ and
$\widetilde{\lambda}$ to the non-coordinate co-frame
\begin{eqnarray}\label{011}
  h\, d\tau = s\,d\Theta - d\widetilde{\lambda}, \nonumber \\
  h\, dy = s\,d\Theta + d\widetilde{\lambda},
\end{eqnarray}
where $\tau$ is proper time in every point of space. 
It is easy to prove that the corresponding derivatives commute between 
themselves,
$$
\left[\partial_{\tau},\,\partial_{y}\right] = 0.
$$
Then Eqs. (\ref{09}) and (\ref{010}) reduce to the form
\begin{equation}\label{012}
    \left\{-\, i\,\hbar \,\partial_{\tau_{c}} - \frac{1}{c}\, 2 \pi^{2} a^{3} 
\rho_{0} \right\} \Psi = 0, \qquad \partial_{y} \Psi = 0,
\end{equation}
where $d\tau_{c} = h\,d\tau$.

From the variation of action with respect to $\widetilde{\lambda}$ it follows 
the conservation law 
\begin{equation}\label{013}
   a^{3} \rho_{0} = \mbox{const} .
\end{equation}
It describes a conserved macroscopic value which characterizes the number of
particles. For example, if a perfect fluid is composed of baryons, then Eq. 
(\ref{013}) describes the conservation of baryon number.

From the variation of action with respect to $\Theta$ it follows that the specific 
entropy is conserved
$$s = \mbox{const}.$$
From Eq. (\ref{08}) one can see that it is convenient to take the matter component  
$\rho$ in the form of relativistic matter (radiation) with the equation of state 
$p = \frac{1}{3} \rho$. In this case
\begin{equation}\label{014}
    a^{4} \rho = \mbox{const}.
\end{equation}
Denoting the constant in Eq. (\ref{014}) as $E$, while in Eq. (\ref{013}) as
$E_{0}$, we obtain the equations
\begin{equation}\label{015}
    \left\{-\, i\,\hbar \,\partial_{\tau_{c}} - \frac{1}{c}\,2 \pi^{2} E_{0} 
\right\} \Psi = 0,
\end{equation} 
\begin{equation}\label{016}
     \left\{-\, \left(\frac{2 G \hbar}{3 \pi c^{3}} \right)^{2}\partial^{2}_{a} + 
a^{2} + \frac{G \hbar^{2}}{3 \pi^{3} c^{2}}\,\, \frac{1}{a^{2}}\,\, 
\partial^{2}_{\phi}\, -\,\frac{8 \pi G}{3 c^{4}}\,\,[a^{4} V(\phi) + E]\right\} \Psi 
= 0,
\end{equation}
where, according to Eq. (\ref{012}), the wave function $\Psi$ does not depend on $y$.

Eq. (\ref{015}) describes the evolution of the state $\Psi$ with respect to the time
variable $\tau_{c}$. Eq. (\ref{016}) does not contain $\tau_{c}$ explicitly.  
At this point there is a close analogy with properties of closed systems
in quantum mechanics.

The constants $E_{0}$ and $E$ are dimensional quantities,  $[E_{0}] = \mbox{energy}$, 
$[E] = \mbox{energy} \times \mbox{length}$. It is convenient to rewrite Eqs.
(\ref{015}) and (\ref{016}) for dimensionless quantities. With that end in view we
bring in correspondence
$$a \rightarrow \frac{a}{l_{P}}, \quad \phi \rightarrow \frac{\phi}{\phi_{P}}, \quad 
\tau_{c} \rightarrow \frac{\tau_{c}}{l_{P}},\quad 
V \rightarrow \frac{V}{\rho_{P}}, \quad E_{0} \rightarrow \frac{4 \pi^{2}}{m_{P} 
c^{2}}\, E_{0}, \quad E \rightarrow \frac{4 \pi^{2}}{\hbar c}\, E,$$
where we have dimensionless quantities from the left, and
$$l_{P} = \sqrt{\frac{2 G \hbar}{3 \pi c^{3}}}, \quad 
\phi_{P} = \sqrt{\frac{3 c^{4}}{8 \pi G}},\quad
t_{P} = \frac{l_{P}}{c}, \quad 
\rho_{P} = \frac{3 c^{4}}{8 \pi G l_{P}^{2}}, \quad
m_{P} = \frac{\hbar}{l_{P} c}$$
are the Planck values of length, scalar field, time, energy density and mass,
respectively. 
Then Eqs. (\ref{015}) and (\ref{016}) in new dimensionless variables take the form 
\begin{equation}\label{017}
    \left\{-\, i\,\partial_{\tau_{c}} - \frac{1}{2}\, E_{0} \right\} \Psi = 0,
\end{equation}
\begin{equation}\label{10}
    \left\{-\,\partial^{2}_{a}  + \frac{2}{a^{2}}\,\, \partial^{2}_{\phi} + a^{2} - 
a^{4} V(\phi) - E\right\} \Psi = 0.
\end{equation}
Eq. (\ref{017}) has a particular solution in the form 
\begin{equation}\label{9}
    \Psi = \mbox{e}^{\,\frac{i}{2}\, E_{0} \tau_{c}} \psi,
\end{equation}
where $\psi$ is a function which depends on $a$ and $\phi$ only and is determined by
Eq. (\ref{10}). If we pass from the time variable $\tau_{c}$ to $\overline{\tau} = 
\frac{E_{0}}{2E}\, \tau_{c}$, as a result we arrive in Eqs. (\ref{10}) and (\ref{9}) 
to an analogy with the Schr\"{o}dinger equation for stationary states. We have 
obtained Eqs. (\ref{10}) and (\ref{9}) previously in Refs. \cite{Kuz1,KuzK} 
within the bounds of the scheme for incorporating of a reference system in general 
relativity through the introduction of a coordinate condition.

Let us note that we can obtain Eqs. (\ref{017}) and (\ref{10}) even
without an introduction of proper time by means of Eqs. (\ref{011}). 
It is possible to build a time 
variable from the matter variables (e.g. Refs. \cite{Lun,Dem}). We can consider e.g. 
$\Theta$ as a time variable \cite{Kuz4}. (On the correspondence 
between the thermasy and proper time see Ref. \cite{Kij}.)

Second order partial differential equation (\ref{10}) is given on the intervals
$0 \leq a < \infty$ and $- \infty < \phi < \infty$. It should be supplemented with
boundary conditions. We suppose that at 
$a \rightarrow 0$ the function $\psi(a, \phi) \rightarrow 
\mbox{const}$. At $a \rightarrow \infty$ its form depends on the properties of the
potential $V(\phi)$. It will determine the behaviour of $\psi$ on the boundaries
$\phi \rightarrow \pm \infty$.

As is well known, the energy density of a uniform scalar field can be
written as
\begin{equation}\label{11}
    \rho_{\phi} = \frac{2}{a^{6}}\,\pi_{\phi}^{2} + V(\phi),
\end{equation}
where we have used an explicit form of the momentum $\pi_{\phi}$ canonically
conjugate with $\phi$ \cite{KuzK},
\begin{equation}\label{12}
    \pi_{\phi} = \frac{1}{2}\, a^{3}\, \frac{d\phi}{d\tau}.
\end{equation}
By analogy we determine the energy density operator for the field $\phi$,
\begin{equation}\label{13}
    \hat{\rho}_{\phi} = -\, \frac{2}{a^{6}}\,\partial_{\phi}^{2} + V(\phi).
\end{equation}
Then the operator of the total energy density in the quantum system under 
consideration will have the form
\begin{equation}\label{14}
    \hat{\rho}_{tot} = \hat{\rho}_{\phi} + \rho,
\end{equation}
where $\rho$ is the same as in Eq. (\ref{014}), and Eq. (\ref{10}) after 
multiplying from the left by $a^{- 4}$ and averaging over the state $\Psi$ 
normalized in one way or another (see below), can be written as
\begin{equation}\label{17}
    \langle \hat{G}_{00} \rangle = 3 \langle \hat{T}_{00}\rangle,
\end{equation}
where the operators are
\begin{equation}\label{18}
    \hat{G}_{00} = \frac{3}{a^{4}}\, \left(\hat{\pi}_{a}^{2} + a^{2} \right), \qquad 
\hat{T}_{00} = \hat{\rho}_{tot}.
\end{equation}
Comparing Eqs. (\ref{17}) and (\ref{18}) with the Einstein equations in general 
relativity we arrive at a conclusion that the operator  
$\hat{G}_{00}$ can be considered as a generalization to quantum theory of the 
correspondent component of the Einstein tensor, while $\hat{T}_{00}$ is the
operator of (00)-component of the energy-momentum tensor (stress tensor) of matter.
The relation (\ref{17}) can be used to find quantum corrections to the 
Einstein-Friedmann equation of classical theory of gravity.

\newpage

\begin{center}
     \textbf{3. Stationary states of the quantum universe}
\end{center}

Let us study the properties of the quantum universe described by the steady-state 
equation (\ref{10}). Since the universe is supposed to be closed, then  
one can introduce a notion of the mass of the universe as a product
of its matter density and a comoving volume. 
In the units under consideration the mass of a scalar field 
in the universe with the scale factor $a$ is equal to
\begin{equation}\label{19}
    M_{\phi} = \frac{1}{2}\,a^{3} \rho_{\phi}.
\end{equation}
This value will be associated with the operator
\begin{equation}\label{20}
    \hat{H}_{\phi} = \frac{1}{2}\,a^{3}\,\hat{\rho}_{\phi},
\end{equation}
where $\hat{\rho}_{\phi}$ is (\ref{13}), while $\frac{1}{2}\, a^{3}$ is a comoving 
volume. Then Eq. (\ref{10}) takes the form
\begin{equation}\label{21}
   \left(-\, \partial_{a}^{2} + a^{2} - 2 a \hat{H}_{\phi} - E\right) \psi = 0.
\end{equation}

Further let us suppose that the potential $V(\phi)$ is a smooth function of $\phi$.
Let there exists a value of the field $\phi = \sigma$ at which the function
$V(\phi)$ has a minimum, while the value $\sigma$ itself corresponds to the true 
vacuum of the field $\phi$, $V(\sigma) = 0$ (an absolute minimum 
\cite{Col}). Then near the point $\phi = \sigma$ the following representation
is valid
\begin{equation}\label{22}
    V(\phi) = \frac{{m}_{\sigma}^{2}}{2}\, (\phi - \sigma)^{2},
\end{equation}
where ${m}_{\sigma}^{2} = [d^{2}V(\phi)/d\phi^{2}]_{\sigma} > 0$. 

Let us make a scaling transformation of the field $\phi$ and introduce a new
variable $x$ which describes a deviation of the field $\phi$
from its vacuum state $\sigma$, 
\begin{equation}\label{23}
    x = \left(\frac{m_{\sigma} a^{3}}2{}\right)^{1/2} (\phi - \sigma).
\end{equation}
The operator (\ref{20}) takes the form
\begin{equation}\label{24}
    \hat{H}_{\phi} = \frac{m_{\sigma}}{2} \left(-\, \partial_{x}^{2} + x^{2}\right).
\end{equation}
Let us introduce the eigenfunctions of harmonic oscillator $u_{k}(x)$ as a
solution of the equation
\begin{equation}\label{25}
    \left(-\, \partial_{x}^{2} + x^{2}\right)\,u_{k}(x) = (2k + 1)\,u_{k}(x),
\end{equation}
where $k = 0,\, 1,\, 2,\, ...$ is the number of the state of the oscillator, 
 $$u_{k}(x) = [\sqrt{\pi} k!\, 2^{k}]^{- 1/2}\, \mbox{e}^{- x^{2}/2} 
H_{k}(x),$$ $H_{k}(x)$ are the Hermitian polynomials. Then we find
\begin{equation}\label{26}
    \hat{H}_{\phi} u_{k} = M_{k} u_{k},
\end{equation}
where
\begin{equation}\label{27}
     M_{k} = m_{\sigma} \left(k + \frac{1}{2}\right).
\end{equation}

The value  $M_{k}$ can be interpreted
as an amount of matter-energy (or mass) in the universe related to a scalar field. 
In the second quantization formalism this energy is represented
in the form of a sum of excitation quanta of the spatially coherent
oscillations of the field $\phi$ about the equilibrium state $\sigma$, $k$ is the
number  of these excitation quanta. Such oscillations correspond to a condensate of 
zero-momentum $\phi$ quanta with the mass $m_{\sigma}$. The mass $m_{\sigma}$ is
determined by the curvature of the potential $V(\phi)$ near $\phi = \sigma$.

Taking into account (\ref{26}) we shall look for the solution of Eq. (\ref{21})
in the form of the superposition of the states with all possible values
of the quantum number $k$ (and, correspondingly, with all possible masses $M_{k}$)
\begin{equation}\label{28}
    \psi = \sum_{k} f_{k}(a)\,u_{k}(x).
\end{equation}
Substituting (\ref{28}) into (\ref{21}) and using the 
orthonormality condition for the
states $u_{k}(x)$, $\langle u_{k}|u_{k'}\rangle = \delta _{k k'}$, we obtain the
equation for $f_{k}(a)$,
\begin{equation}\label{29}
     \left(-\, \partial_{a}^{2} + a^{2} - 2 a M_{k} - E\right) f_{k}(a) = 0.
\end{equation}
This equation has an analytical solution decreasing at $a \rightarrow \infty$ which
has the form of the wave function of an oscillator perturbed by the mass term 
$- 2 a M_{k}$,
\begin{equation}\label{30}
    f_{k}(a) \equiv f_{n,k}(a) = N_{n,k}\, \mbox{e}^{-\, \frac{1}{2}(a - M_{k})^{2}} 
H_{n} (a - M_{k})
\end{equation}
at
\begin{equation}\label{31}
    E \equiv E_{n,k} = 2n + 1 - M_{k}^{2},
\end{equation}
where $n = 0,\,1,\,2, ...$ is the number of the state of the 
quantum universe at a given 
$k$-state of a condensate (with the mass $M_{k}$) in the potential well
\begin{equation}\label{390}
    U(a) = a^{2} - 2 a M_{k}.
\end{equation}
For the states $f_{n,k}(a)$, 
normalized by the condition $\langle f_{n,k}|f_{n,k}\rangle = 1$, the
normalization factor $N_{n,k}$ is equal to
\begin{eqnarray}\label{32}
  N_{n,k} & = & \left\{2^{n-1} n! \sqrt{\pi}\, [\mbox{erf}\,M_{k} + 1] \right. 
\nonumber \\
& & \left. -\,\mbox{e}^{- M_{k}^{2}}\, \sum _{l=0}^{n-1}\, \frac{2^{l}n!}{(n-l)!}\, 
H_{n-l}(M_{k})\,H_{n-l-1}(M_{k})\right\}^{- \frac{1}{2}},
\end{eqnarray}
where
$$\mbox{erf}\,M = \frac{2}{\sqrt{\pi}}\, \int_{0}^{M}\! dt\,\mbox{e}^{- t^{2}}$$
is a probability integral. From the properties of the function $\mbox{erf}\,M$
and the properties of the Hermitian polynomials $H_{n}(M)$ it follows that the 
normalization factor 
(\ref{32}) for $M_{k} > 1$ is equal to 
\begin{eqnarray}\label{34}
  N_{n,k} = \left\{2^{n} n! \sqrt{\pi} - O\left((2 M_{k})^{2n-1}\,\mbox{e}^ {- M_{k}^{2}} \right)\right\}^{- \frac{1}{2}}.
\end{eqnarray}
One can neglect the exponential addition for the states of the quantum universe with
the large enough masses $M_{k}$ of a condensate. In this case $N_{n,k}$ does
not depend on $k$, and its numerical value coincides with the numerical value of
the normalization factor of the wave function of ordinary harmonic oscillator in the 
state $n$.

According to (\ref{30}) and (\ref{31}) the quantum states of the universe are
characterized by two quantum numbers $n$ and $k$. The mean value of the scale factor
$a$ in the state (\ref{30}),
\begin{equation}\label{35}
    \bar{a} = \langle f_{n,k}|a|f_{n,k} \rangle,
\end{equation}
equals to
\begin{equation}\label{36}
     \bar{a} = M_{k} + \bar{\xi},
\end{equation}
where
\begin{eqnarray}\label{37}
    \bar{\xi} = N_{n,k}^{2}\, 2^{n-1} n!\, \mbox{e}^{- M_{k}^{2}} \left\{1 + 
\sum_{l=0}^{n-1}\, \frac{2^{l-n}}{(n-l)!}\,H_{n-l}(M_{k})\,H_{n-l-1}(M_{k}) \right\}
\nonumber \\
= \frac{N_{n,k}^{2}}{2}\, n! \, \mbox{e}^{- M_{k}^{2}} \left\{2^{n} + 
O\left((2M_{k})^{2n-1}\right)\right\}.
\end{eqnarray}
According to (\ref{36}) and (\ref{37}) the universe with the Planck mass of
a condensate, $M_{k} = 1$, in the ground (vacuum) state,  $n = 0$, is characterized
by the mean value $\bar{a}_{n=0} = 1.11$ which  coincides with the Planck length by 
an order of magnitude. For the states with $M_{k} > 1$ the mean value (\ref{36})
does not depend on $n$ to within a small summand
$\sim O((2M_{k})^{2n-1}\,\mbox{e}^{- M_{k}^{2}})$ and is determined by the mass
$M_{k}$ only,
\begin{equation}\label{39}
    \bar{a} = M_{k}\qquad \mbox{at}\qquad M_{k}\gg 1.
\end{equation}
The mass $M_{k}$ determines also the value of an absolute minimum of the effective 
potential $U(a)$ (\ref{390})
of Eq. (\ref{29}), $U(M_{k}) = -\, M_{k}^{2}$.

\begin{center}
  \textbf{4. The nucleation rate of the universe from the initial 
cosmological singularity point}  
\end{center}

The wave function $f_{n,k}(a)$ describes the universe in the state with the 
quantum numbers $n$ and $k$ and depends on the scale factor $a$. 
The ground state of the quantum universe is characterized by the Planck parameters.
The wave function $f_{n,k}(a)$ itself can be considered at the point $a = 0$. 
According to quantum field theory a particle decay 
rate is determined by the expression
\begin{equation}\label{40}
    \Gamma_{\psi} = \overline{v\,\sigma}_{r}\,|\psi(0)|^{2},
\end{equation}
where $v$ is the relative velocity of decay products, $\sigma_{r}$ is the reaction 
cross-section, and a bar means an averaging over non-recording parameters (e.g., 
over initial spin states), $\psi(0)$ is the wave function of a particle 
before the decay 
in the origin (at zero distance). In accordance with that the value
\begin{equation}\label{41}
    \Gamma_{n,k} = \overline{v\,\sigma}_{r}\,|f_{n,k}(0)|^{2},
\end{equation}
can be interpreted as a rate of nucleation of the universe in the $n,k$-state
from the initial cosmological singularity point $a = 0$. In accordance with classical 
view the nucleation of the universe is the process in time with an expansion 
initiation at some instant $\tau$ which it is convenient to choose as $\tau = 0$.
The velocity $v$ can be naturally identified with the expansion rate 
$v = \frac{da}{d\tau}$ at $\tau = 0$, 
where $\tau$ is some time parameter which ensures
the boundary condition $a(\tau = 0) = 0$. The cross-section $\sigma_{r}$ will be set
equal to $\sigma_{r} = \pi\, a^{2}$. The dependence  $a(\tau )$ can be found from the 
condition of finiteness of $\overline{v\,\sigma}_{r}$ at the point $a = 0$. We put
\begin{equation}\label{42}
   \overline{v\,\sigma}_{r}\, \equiv \lim_{a \rightarrow 0}\,\left(\frac{da}{d\tau}\,\pi \,a^{2}\right) = \mbox{const}.
\end{equation}
It is easy to make sure that at $\mbox{const} \neq 0$ the single possible case is 
$a(\tau ) \sim \tau ^{1/3}$ which is realized when primordial matter is
described by the extremely rigid equation of state at the point $a(\tau = 0) = 0$,
$$
    p_{in} = \rho_{in},
$$
where $p_{in}$ and $\rho_{in}$ are the pressure and energy density of a primordial 
scalar field at the moment of nucleation of the universe. From the point of view of 
semi-classical approximation such an equation of state is realized in the primordial 
scalar field $\phi$ taken in the state of its true vacuum $\phi = \sigma$, 
where $V(\sigma) = 0$,
\begin{equation}\label{43}
    \rho_{\sigma} = \frac{1}{2}\, \left(\frac{d\phi}{d\tau}\right)_{\sigma}^{2} = 
p_{\sigma},
\end{equation}
when all energy of the field $\phi$ is concentrated in its kinetic part. Such a 
state is unstable and should turn into the state with a condensate of $\phi$ quanta.
This transition will look like a nucleation of the universe from the point $a = 0$
with the extremely rigid equation of state (\ref{43}) 
and the wave function $f_{n,k}(0)$, the
square of which determines the rate of such a process in accordance with (\ref{41}).
(Physics of transition from $a = 0$ to $\bar{a} \neq 0$ is considered below at
the end of this section.)

Using an explicit form of the function $f_{n,k}(a)$ (\ref{30}) and keeping the main
term in $N_{n,k}$ (\ref{34}) only, we find
\begin{equation}\label{44}
    \Gamma_{n,k} \simeq \mbox{const}\, \frac{2^{n}}{\sqrt{\pi}}\, P(n),
\end{equation}
where 
$$P(n) = \frac{\langle n \rangle^{n}}{n!}\, \mbox{e}^{- \langle n \rangle}$$
is the Poisson distribution with the mean value 
$\langle n \rangle = M_{k}^{2}$ of the quantum number $n$. 
The total nucleation rate of the universe $ \Gamma = \sum_{n}\, \Gamma_{n,k}$ 
is given by the mean value $\langle 2^{n} \rangle$ over the Poisson distribution.
It appears to be exponentially high
\begin{equation}\label{45}
    \Gamma \simeq  \frac{\mbox{const}}{\sqrt{\pi}}\, \mbox{exp} \{M_{k}^{2}\}.
\end{equation}

According to quantum theory  
the square of modulus of the wave function at
the origin determines the particle number density at this point. In the case
of the quantum universe
\begin{equation}\label{47}
    |f_{n,k}(0)|^{2} \sim L_{n,k}^{- 3},
\end{equation}
where $L_{n,k}$ is the linear dimension of the region from which the universe
nucleates. Hence we find that
\begin{equation}\label{48}
    L_{n,k} \sim \left( \frac{\sqrt{\pi} n!}{2^{n} M_{k}^{2}}\right)^{1/3} 
\mbox{exp}\{\frac{M_{k}^{2}}{3}\}.
\end{equation}

For $n = 0$ and $M_{k} = 1$ we have $L_{0,k} \sim 1.69$. This value is consistent 
with the mean value $\bar{a}_{n=0} = 1.11$ calculated above for the 
universe with the Planck mass of a condensate.  

Given estimations show that the universe nucleates with a finite rate into the
state with the Planck parameters (mass and spatial dimensions). The larger masses
of a condensate of nucleating universe correspond to the larger primordial 
dimen\-sions. The total nucleation rate of the universe  
obeys an exponential law and corresponds to an ``explosion'' from the point $a = 
0$. It can be identifies with zero time of the Big Bang. 

As we have shown in Ref. \cite{KK2} a condensate of $\phi$ quanta with the total mass
$M_{k} \neq 0$ has an antigravitating property and is described by the vacuum 
equation of state $p_{k} = -\, \rho_{k}$,
where $\rho_{k} = \frac{2 M_{k}}{\bar{a}^{3}}$.  
The universe from an unstable state with the extremely rigid
equation of state at the point $a = 0$ passes into the ground state with the Planck
mass of a condensate and the Planck scale factor. 
As soon as the mass of a condensate reaches nonzero values the equation of state of a
primordial scalar field changes 
from the extremely rigid equation of state to the vacuum one and a condensate
acquires an antigravitating property.
The growth of $M_{k}$ leads to the growth of 
antigravitation and as a 
consequence triggers a subsequent growth of $\bar{a}$ of the quantum 
universe which at that undergoes an accelerating expansion. 

\begin{center}
    \textbf{5. Probabilities of transitions between the states of the 
universe with different masses of a condensate}
\end{center}

According to (\ref{28}), (\ref{30}) the function $f_{n,k}(a)$ can be interpreted as a 
state vector which describes the universe in the $n$-th state with the mass
of a condensate $M_{k}$. The states $f_{n,k}$ are orthogonal with respect to the
quantum number $n$ up to small terms $\sim \exp\{-M_{k}^{2}\}$. 
If the states $f_{n,k}$ and $f_{n',k'}$ correspond to different
masses, $M_{k} \neq M_{k'}$, then they are eigenfunctions of different operators
and therefore, generally speaking, are nonorthogonal between themselves. The
correspondent overlap integral is $\langle f_{n,k}|f_{n',k'}\rangle \neq 0$. The
evolution of the universe takes place is such a way that it passes from, say, the 
$n,k$-state in one potential well $U(a)$ (\ref{390}) into the state with the quantum
numbers $n',k'$ in another well, where the number of the state $n'$ may differ from
$n$ or be equal to it, but an index $k'$ that numbers a quantity of $\phi$ quanta
differs from $k$.

Eq. (\ref{29}) describes the stationary states. Therefore when calculating the
transition probability $w(n,k \rightarrow n',k')$ we use the model of the 
instantaneous change of the state of quantum system. Then
\begin{equation}\label{49}
    w(n,k \rightarrow n',k') = |\langle f_{n',k'}|f_{n,k}\rangle |^{2}.
\end{equation}
It is easy to show that the normalization condition
\begin{equation}\label{50}
    \sum_{n'}\, w(n,k \rightarrow n',k') = 1
\end{equation}
is satisfied to within exponentially small terms. 

Using the explicit form of the function  $f_{n,k}(a)$ (\ref{30}), 
integrating by parts, and neglecting the terms  $\sim \mbox{exp}\{-\,M_{k}^{2}\}$ for $M_{k} \gg 1$ we find
\begin{equation}\label{54}
    \langle f_{n',k'}|f_{n,k}\rangle = 2\,\sqrt{\pi}\, N_{n',k'} 
N_{n,k}\,\xi_{0}^{n-n'}\,\mbox{e}^{-\,\xi_{0}^{2}}\,\sum_{i=0}^{n'}\,(-\,1)^{i}\,
\frac{2^{n'-i}\,n'! \,n!\,\xi_{0}^{2i}}{i!(n'-i)!(n-n'+i)!}
\end{equation}
for $n \neq 0$ and 
\begin{equation}\label{55}
    \langle f_{n',k'}|f_{0,k}\rangle = 2\,\sqrt{\pi}\, N_{n',k'} 
N_{0,k}\,\xi_{0}^{n'}\,\mbox{e}^{-\,\frac{1}{4}\xi_{0}^{2}},
\end{equation}
where we denote $\xi_{0} = M_{k'} - M_{k}$. 
The transition probability (\ref{49}) at $n' > n \geq 1$ is equal to
\begin{equation}\label{56}
    w(n,k \rightarrow n',k') \simeq \frac{1}{2^{n'-n-2}}\, 
\frac{n'!}{n![(n'-n)!]^{2}}\, \xi_{0}^{2(n'-n)}\,\mbox{e}^{-2 \xi_{0}^{2}},
\end{equation}
where we have used the expression (\ref{54}) and keep only the main term with
$i = n' - n$. The probability of transition between the states with $n' < n$
follows from (\ref{56}) after the substitution $n \leftrightarrow n'$ and it can
be obtained from (\ref{54}) keeping the main term with $i = 0$.

According to (\ref{56}) when the difference $\xi_{0} = m_{\sigma}(k' - k)$ grows
the transition probability falls almost exponentially. For the $\phi$ quanta
with the masses $m_{\sigma} \sim 1$ the transitions between
the states of a condensate with the close numbers $k$ and $k'$ in different
potential wells $U(a)$ have the highest probability. For the mass
$m_{\sigma} \sim 10^{-19}\,(\sim 1\,\mbox{GeV})$ the parameter $\xi_{0} \sim 1$
for the difference $(k'-k) \sim 10^{19}$.

Nonzero transition probability points to a possibility in principle for the universe
to evolve as a result of transitions between quantum states. An increase (decrease)
in the mass of a condensate means 
an increase (decrease) in the mean value of the scale
factor of the universe. From the point of view of semi-classical theory the universe
will expand (contract).

Using Eq. (\ref{55}) one can calculate the probability of transition of the 
universe from the ground (vacuum) state, $n = 0$, to any other state. It obeys
the Poisson distribution
\begin{equation}\label{57}
    w(0,k \rightarrow n',k') = \frac{\langle n' \rangle^{n'}}{n'!}\, 
\mbox{e}^{- \langle n' \rangle}
\end{equation} 
with the mean value $\langle n' \rangle =\xi_{0}^{2}/2$ of the 
quantum number $n'$. If
$\langle n' \rangle \ll 1$, then this probability is small 
and it decreases rapidly with
an increase in $n'$. The highest probability in this case has the transition
$0,k \rightarrow 1,k'$, 
\begin{equation}\label{58}
     w(0,k \rightarrow 1,k') = \langle n' \rangle - O(\langle n' \rangle^{2}).
\end{equation}
The transition vacuum $\rightarrow$ vacuum from different potential wells $U(a)$
is given by an exponent
\begin{equation}\label{59}
    w(0,k \rightarrow 0,k') = \mbox{e}^{- \langle n' \rangle}.
\end{equation}
It means that the transitions from vacuum to non-vacuum states occur with the
overwhelming probability. The total probability of such transitions equals to
\begin{equation}\label{60}
    w(\mbox{vac}\rightarrow \mbox{nonvac}) \equiv \sum_{n'\neq 0}\,w(0,k \rightarrow 
n',k') = 1 - \mbox{e}^{- \langle n' \rangle}.
\end{equation}
The ratio of the total transition 
probability (\ref{50}) at $n = 0$ to probabilities of 
transitions between the vacuum states is equal to
\begin{equation}\label{61}
    \frac{\sum_{n'}\,w(0,k \rightarrow n',k')}{w(0,k \rightarrow 
0,k')} = \mbox{e}^{\langle n' \rangle},
\end{equation}
i.e. against a background of vacuum-vacuum transitions the probabilities of 
transitions into non-vacuum states look like exponentially high.

\begin{center}
     \textbf{6. Conclusion}
\end{center}

In this paper we calculate the spectrum of stationary states 
(\ref{31}) and the wave functions (\ref{28}), (\ref{30}) 
of the homogeneous and isotropic 
universe in 
the epoch of matter-energy production from a primordial uniform scalar field on 
basis of the exact solution of Eq. (\ref{10}) of quantum model. 
Produced matter-energy
represents itself a condensate of excitation quanta of oscillations of a scalar field
above its true
vacuum state. The mass of a condensate (\ref{27}) and the mean value of the scale
factor (\ref{35}) in a given state of the universe 
are connected between themselves by 
the linear expression (\ref{36}). 
Let us note that the condition (\ref{39}) is a mathematical formulation of the Mach's
principle proposed by Sciama \cite{Dic} (see also \cite{KK3}).
The universe in an arbitrary state (\ref{28})
is described by the superposition of the states with all possible masses of a 
condensate. The nucleation rate of the universe from the initial cosmological
singularity point (\ref{44}) appears to be non-zero, while the total
nucleation rate obeys the exponential (explosive) law (\ref{45}). The nucleation
of the universe takes place as a result of its transition from the initial 
cosmological singularity point with the extremely rigid equation of state of a
primordial scalar field into the state with the non-zero mass of a condensate with 
the vacuum equation of state. The universe being nucleated in the ground (vacuum)
state has the Planck parameters. The evolution of the universe is described as 
transitions with the non-zero probability (\ref{56}) between the states of the 
universe with different masses of a condensate. An increase (decrease) in this mass
leads to an expansion (contraction) of the universe. 

The classical universe can be imagined as a ball which is situated in the minimum
of the effective potential $U(a)$ (\ref{390}) and moves together with this minimum
in the direction of large $a$ with grows of its mass. From the point of view of
classical theory such a motion (expansion of the universe) 
can continue for an arbitrary
long period of time. System has no restrictions in dimensions and mass.

\end{document}